\documentclass[reqno,11pt]{amsart}
\usepackage{amsmath,amssymb,amsthm,graphicx,a4wide,enumerate,url}
\usepackage[small,bf]{caption}
\usepackage{enumitem,subfigure}

\usepackage{scalerel}
\usepackage{tikz}
\usetikzlibrary{svg.path}
\definecolor{orcidlogocol}{HTML}{A6CE39}
\tikzset{
  orcidlogo/.pic={
    \fill[orcidlogocol] svg{M256,128c0,70.7-57.3,128-128,128C57.3,256,0,198.7,0,128C0,57.3,57.3,0,128,0C198.7,0,256,57.3,256,128z};
    \fill[white] svg{M86.3,186.2H70.9V79.1h15.4v48.4V186.2z}
                 svg{M108.9,79.1h41.6c39.6,0,57,28.3,57,53.6c0,27.5-21.5,53.6-56.8,53.6h-41.8V79.1z M124.3,172.4h24.5c34.9,0,42.9-26.5,42.9-39.7c0-21.5-13.7-39.7-43.7-39.7h-23.7V172.4z}
                 svg{M88.7,56.8c0,5.5-4.5,10.1-10.1,10.1c-5.6,0-10.1-4.6-10.1-10.1c0-5.6,4.5-10.1,10.1-10.1C84.2,46.7,88.7,51.3,88.7,56.8z};
  }
}

\newcommand\orcidicon[1]{\href{https://orcid.org/#1}{\mbox{\scalerel*{
\begin{tikzpicture}[yscale=-1,transform shape]
\pic{orcidlogo};
\end{tikzpicture}
}{|}}}}

\usepackage{hyperref} 

\usepackage{graphics}
\graphicspath{ {./figures/} }
\theoremstyle{plain}

\theoremstyle{definition}

\theoremstyle{remark}

\newcommand{\citep}[1]{\cite{#1}}

\usepackage{color}

\begin{document}

\title{Macroscopic Manifestations of Traffic Waves in Microscopic Models}
\author{Nour Khoudari, Rabie Ramadan, Megan Ross, Benjamin Seibold}
\date{\today}

\begin{abstract}
Traffic waves can rise even from single lane car-following behaviour. To better understand and mitigate traffic waves, it is necessary to use analytical tools like mathematical models, data analysis, and micro-simulations that can capture the dynamics of real traffic flow. In this study, we isolate car-following dynamics and present a systematic hierarchy of tests that connect the microscopic scale with the meaningful macroscopic effective state in the presence of waves. This allows insights with precise attributable cause-to-effect relationships of specific observed traffic patterns. We establish a principled way of generating macroscopic flow quantities from microscopic models in the unstable regime. Those quantities are then used to study how the corresponding non-equilibrium wave structures manifest in the fundamental diagram, based on three basic scenarios that can serve as building blocks for understanding more complex micro-simulation studies. Finally, this study gives insight on the shapes of the reduced fundamental diagrams for different commonly used microscopic models.
\end{abstract}

\maketitle

\section{Introduction}
Vehicular traffic flow is an intrinsic multiscale system, where the nonlinear vehicle-scale dynamics result in emergent structures on larger scales, and thus affect the temporal evolution of the macroscopic flow properties such as vehicle density and bulk flow rates. And in turn, macroscopic properties such as average in-/out-flow rates at domain boundaries or fundamental diagrams ultimately cascade down to the micro-scale vehicle dynamics. Macroscopic descriptions of traffic flow, such as the Lighthill-Whitham-Richards model \citep{LighthillWhitham1955} or many more complex models \citep{AwRascle2000, Zhang2002, LebacqueMammarHajSalem2007}, describe the evolution of the macroscopic flow quantities via fundamental conservation principles combined with empirical laws fitted to macro-scale measurements (and thus completely circumvent the modelling of the vehicle-scale dynamics). In turn, micro-simulators \citep{HelbingHenneckeShvetsovTreiber2002} numerically resolve precisely those vehicle-vehicle interactions, and relevant macroscopic observables can then be extracted post-hoc from ensemble simulations. Those macro-scale quantities are generally viewed as empirical outcomes that can be measured from the micro-simulation, rather than resulting from systematic mathematical principles (except in particularly simple situations, see below).

A key property of real-world traffic is that it can dynamically exhibit non-equilibrium flow features, such as phantom traffic jams and traffic waves. These features have been experimentally demonstrated in multiple cases \citep{SugiyamaFukuiKikuchiHasebeNakayamaNishinariTadakiYukawa2008, SternCuiDelleMonacheBhadaniBuntingChurchillHamiltonHaulcyPohlmannWuPiccoliSeiboldSprinkleWork2018}, and they can be interpreted as instabilities and nonlinear travelling waves in traffic models, both microscopic \citep{Pipes1953, Newell1961, BandoHesebeNakayama1995} and macroscopic \citep{Payne1971, Payne1979, AwRascle2000, Greenberg2004, FlynnKasimovNaveRosalesSeibold2009}.\footnote{There are also other ways to describe traffic, such as kinetic, cellular, etc.~\citep{Daganzo1994, Phillips1979, RamadanRosalesSeibold2019}. The study conducted herein distinguishes two key situations: ``microscopic'' means that the actual vehicle dynamics are resolved; while ``macroscopic'' means that vehicle density and flow rates are studied on scales that average over the vehicle scale.} As a general principle, first-order models that describe the vehicle velocity based on the vehicle positions (without delay) tend to not produce instabilities; in turn, second-order models that describe vehicle (or flow) accelerations based on vehicle positions and velocities, \emph{can} produce instabilities and traffic waves \citep{VANWAGENINGENKESSELS2015445}. Key examples of microscopic car-following models that fall into this latter category are the Optimal Velocity Model (OVM), discussed in \S\ref{subsubsec:OVM}, 
and the Intelligent Driver Model (IDM), discussed in \S\ref{subsubsec:IDM}.
The fact that those models (in suitable parameter regimes) can produce unstable equilibrium flow and generate traffic waves is well known \citep{BandoHesebeNakayama1995, TreiberHenneckeHelbing2000, wilson2011}. However, what is not established is a systematic, principled way to generate macroscopic flow quantities from those microscopic models in the unstable regime. Specifically, traffic waves in these models manifest as concrete non-equilibrium structures in the fundamental diagram (FD); and this work establishes those structures. It should be noted however that this work does not study the question of finding the macroscopic limits of those microscopic models.

Besides connecting the micro-scale (vehicle dynamics) to the macro-scale (non-equilibrium FD), this work also establishes a systematic procedure how to analyse the emergent structures arising from unstable car-following dynamics, based on the hierarchy of simple scenarios that can serve as building blocks for understanding more complex micro-simulation studies. This procedure provides some fundamental insights on the behaviour of the different car-following models (here: OVM and IDM) in the non-equilibrium flow regime, and the shapes of the corresponding reduced fundamental diagrams.

This manuscript is organised as follows. Section~\ref{sec:prior_work_models} describes the existing traffic models used in this study, and presents some technical extensions thereof. The methods used in this work are presented in \S\ref{sec:methodologies}, specifically: the numerical methods, simulation setups, and techniques used to derive macroscopic fields and interpret traffic waves on the FD. The fundamental results, and discussion thereof, are presented in \S\ref{sec:results_and_discussion}, and conclusions and an outlook of future research directions given in \S\ref{sec:conclusions_outlook}.

\section{Prior Work and Models}
\label{sec:prior_work_models}
This section contains a literature review of microscopic and macroscopic traffic models relevant to this work. For microscopic models, we introduce and discuss the Intelligent Driver Model (IDM), and the Optimal Velocity Model (OVM) with a small twist that we add to the original model. For macroscopic models, we outline the Lighthill-Whitham–Richards (LWR) model. The section is concluded with a discussion of the numerical methods used to efficiently solve the above models.

\subsection{Microscopic Car-Following Models}
\label{subsec:micro-car-following-models}
Microscopic car-following traffic models are systems of ordinary differential equations (ODEs), where the dynamics of each vehicle are described by a dynamic equation of motion. To derive the trajectories of individual vehicles, the ODEs could prescribe the vehicle velocity (first-order models), or the velocity and the acceleration (second-order models). Second-order car-following models are of the form
\begin{equation}
    \label{eq:general-car-following-models}
     \dot{v}(t) = f(s(t), v(t), \Delta v(t))\;.
\end{equation}
Here $s$ is the gap (measured in metres) to the vehicle ahead (the ``lead vehicle''), $v$ is the vehicle's velocity (measured in $\text{m}/\text{s}$), and $\Delta v$ is the relative velocity between the lead vehicle and the vehicle itself, defined for vehicle $i$ as $\Delta v_i = v_{i-1}(t)-v_i(t)$, where $v_i$ is its own speed and $v_{i-1}$ the lead vehicle's speed. Many second order models have been proposed; below we discuss two important examples of them.

\subsubsection{The Intelligent Driver Model (IDM)}
\label{subsubsec:IDM}
The IDM \citep{TreiberHenneckeHelbing2000, HelbingHenneckeShvetsovTreiber2002} is a special case of \eqref{eq:general-car-following-models}, and reads as:
\begin{equation}
    \label{eq:IDM_Equation}
    f(s,v,\Delta v)_{\text{IDM}} = 
    \begin{cases} 
    0 & \text{if}~ v=0~ \text{and}~ \hat{f}\leq 0 \\
    \hat{f} & \text{otherwise}\;,
    \end{cases}
\end{equation}
where $\hat{a}$ is defined as
\begin{equation}
    \hat{f} = a\left[1-\left(\frac{v}{v_{0}}\right)^{\delta}-\left(\frac{s^{*}\left(v,\Delta v\right)}{s}\right)^{2}\right],
\end{equation}
and $s^{*}\left(v,\Delta v\right)$ is defined as
\begin{equation}\label{eq:IDM_Spacing_Equation}
    s^{*}\left(v,\Delta v\right) = s_{0}+vT-\frac{v\Delta v}{2\sqrt{ab}}\;.
\end{equation}
Note that here $\Delta v$ is defined the negative of how it is defined in \citep{TreiberHenneckeHelbing2000}, hence the corresponding term in \eqref{eq:IDM_Spacing_Equation} appears with a minus sign.
The IDM has six parameters: $a$, $b$, $v_0$, $T$, $\delta$, and $s_0$. The parameter $v_0 > 0$ represents the desired velocity on an empty road (measured in $\text{m}/\text{s}$), and $s_0 > 0$ represents the desired minimum spacing between vehicles (measured in $\text{m}$). Moreover, $T>0$ is the desired time headway (the minimum possible time to reach the vehicle ahead, measured in $\text{s}$), and $\delta$ is the acceleration exponent (dimensionless), frequently set to $\delta = 4$ \citep{TreiberHenneckeHelbing2000}. The parameters $a$ and $b$ (commonly measured in $\text{m}/\text{s}^2$) are both positive, and they correspond to the maximum vehicle acceleration and minimum desired comfortable deceleration, respectively.

\subsubsection{The Optimal Velocity Model (OVM)}
\label{subsubsec:OVM}
The OVM, introduced in \citep{BandoHesebeNakayama1995}, is another special case of \eqref{eq:general-car-following-models}, and it reads as:
\begin{equation}
    \label{eq:OVM_Equation}
    f(s,v,\Delta v)_{\text{OVM}} =\alpha\left[V(s)- v\right].
\end{equation}
Here $V(s)$ denotes the optimal velocity function determined by the gap to the vehicle ahead. The optimal velocity function should satisfy the following conditions \citep{BandoHesebeNakayama1995}: monotone increasing, continuous, non-negative, with lower and upper limit boundary (asymptotic at speed limits). Given $V(s)$, the OVM has one free parameter: the sensitivity $\alpha > 0$ (measured in $1/\text{s}$). This model assumes that each vehicle maintains the maximum speed given a large distance to the vehicle ahead, and otherwise aims to adjust its spacing so that its optimal velocity matches the leader's speed.

An important drawback of the model \eqref{eq:OVM_Equation} is the possibility of the crossing of trajectories, representing car collisions. To avoid that, the OVM is augmented with a follow-the-leader term as in \citep{AwKlarMaterneRascle2002}. This term models additional braking and was introduced in the models \citep{GazisHermanRothery1961, HermanPrigogine1971}. The augmented OVM reads as
\begin{equation}
    \label{eq:OVM_FtL}
    f(s,v,\Delta v)_{\text{OVM-FtL}} = \alpha\left[V(s)- v\right]+ \beta\left
    [\frac{\Delta v}{s^\nu}\right],
\end{equation}
where $\nu$ is a positive exponent affecting the range of the impact of velocity calibration to the velocity of the vehicle ahead, and $\beta$ is a positive braking coefficient (measured in $\text{m}^\nu/\text{s}$).

The model \eqref{eq:OVM_FtL} with the augmented follow-the-leader term has another drawback of potentially producing extreme acceleration and deceleration values when the lead vehicle is having a vastly different speed. For example, if a vehicle falls very far behind the vehicle ahead of it, then the corresponding $V(s)$ would approach the speed limit, and in the cases where the velocity of the vehicle $v$ is very small (e.g., due to strong waves), the model \eqref{eq:OVM_FtL} could yield unrealistically strong instantaneous accelerations. 

Hence, we modify the OVM to ensure that it is more realistic and comparable to the IDM in the cases when waves develop. To make the OVM acceleration function more realistic, we limit the possible acceleration and deceleration values by applying a saturation function to the difference in velocity that reads as
\begin{equation}
\label{eq:saturation_function}
    g(u) = \left[\frac{a_\text{m}-b_\text{m}}{2}\right]+\left[\frac{a_\text{m}+b_\text{m}}{2}\right]\tanh{\left[cu-u_0\right]}\;,
\end{equation}
where 
\begin{equation*}
u_0 = \text{atanh}\left[\frac{a_\text{m}-b_\text{m}}{a_\text{m}+b_\text{m}}\right]
\quad\text{and}\quad
c = \frac{2\alpha}{(a_\text{m}+b_\text{m})\text{sech} (u_0^2)}\;.
\end{equation*}
Here $\alpha > 0$ is a sensitivity constant (measured in $1/\text{s}$), $a_\text{m} > 0$ is the maximum acceleration value and $b_\text{m} > 0$ is the maximum deceleration value (both measured in $\text{m}/\text{s}^2$). The function $g$ smoothly transitions from $\lim_{u\to -\infty} g(u) = -b_\text{m}$ through $g(0) = 0$ to $\lim_{u\to +\infty} g(u) = a_\text{m}$ such that $g'(0) = \alpha$. Because of that last property, the saturation function naturally generalises the multiplication by $\alpha$ in \eqref{eq:OVM_FtL}.

The saturation function \eqref{eq:saturation_function} is applied only to the car-following term in \eqref{eq:OVM_Equation}. The follow-the-leader remains uncapped, because that term was added to obtain a model devoid of collisions under all circumstances.
The resulting modified version of \eqref{eq:OVM_FtL} with the saturation function applied, now reads as
\begin{equation}
    \label{eq:OVM_modified}
    f(s,v,\Delta v)_{\text{OVM-Modified}} = g\left[V(s)- v\right]+ \beta\left
    [\frac{\Delta v}{s^\nu}\right]\;.
\end{equation}
Any usage of, and reference to, the OVM in the simulations below corresponds to \eqref{eq:OVM_modified}.

\subsection{Macroscopic Traffic Models}
\label{subsec:macroscopic-models}
Macroscopic traffic models come in the form of partial differential equations (PDEs) that describe aggregate quantities such as traffic density or flow rate, instead of resolving the individual vehicles. A fundamental example is the Lighthill-Whitham-Richards (LWR) model. It is a first-order hyperbolic conservation law, describing the density of vehicles on a road with no entries or exits. It was first proposed in \citep{LighthillWhitham1955, Richards1956} as an equation describing the relation between density $\rho$ and flow rate $q$ as 
\begin{equation}
    \label{LWR-continuity-eq}
    \frac{\partial{\rho}}{\partial t}+\frac{\partial{q}}{\partial x} = 0\;.
\end{equation}
Equation \eqref{LWR-continuity-eq} is often referred to as the continuity equation, and will be used in later sections to denote a FD relationship $q=Q(\rho)$, via a bulk velocity $u$, where $q = \rho u$. The scalar conservation law \eqref{LWR-continuity-eq} satisfies a maximum principle, hence it is devoid of dynamic instabilities and cannot be used as a model for phantom traffic jams. 
It is important to note that other macroscopic models exist that allow the study of phantom traffic jams like the inhomogeneous Aw-Rascle-Zhang (ARZ) model \citep{AwRascle2000, Zhang2002, Greenberg2004}, which provides a second evolution equation for the bulk velocity field and can be seen as a special case of generic macroscopic second-order models \citep{LebacqueMammarHajSalem2007, FanHertySeibold2014}. Note that second-order macroscopic models are not considered here as this work only studies the manifestations of microscopic models, but not their macroscopic limits.

\section{Methodologies}
\label{sec:methodologies}
In this section we establish the methodologies how to fundamentally interpret microscopic models with instabilities and noise, and how to extract macroscopic quantities from microscopic models. A detailed description of how to characterise waves is provided, and the section is concluded by a comparison between waves in the OVM vs.~waves in the IDM.

\subsection{Simulation Setups}
\label{subsec:setups}
In this work, the setups used for the traffic simulations can be categorised into three main types: ring roads, infinite roads, and bottlenecks. For all setups, we consider vehicles on single-lane roads with positions $x_i(t)$ and speeds $v_i(t)$, where vehicle $i$ follows vehicle $(i-1)$ and the equations of motion of all vehicles follow \eqref{eq:general-car-following-models}. In a ring road setup, $N$ vehicles circulate in a loop, where the $1^{\text{st}}$ vehicle follows the $N^{\text{th}}$. Even though a ring road is not a realistic representation of traffic on highways, this setup creates a controlled environment to study congestion, traffic waves, and car-following behaviour, as for example employed in real-world experiments \citep{SugiyamaFukuiKikuchiHasebeNakayamaNishinariTadakiYukawa2008, Tadakietal2013, SternCuiDelleMonacheBhadaniBuntingChurchillHamiltonHaulcyPohlmannWuPiccoliSeiboldSprinkleWork2018}. The infinite road setup simulates traffic on an unbounded road providing a realistic representation of traffic on real-world highways when a lead vehicle with a large gap ahead is present. Whereas, the bottleneck road setup simulates traffic on a bounded road segment with prescribed inflow and outflow conditions, where the outflow condition often causes localised disruption in traffic and growing traffic jams. In real life, a bottleneck can be caused by the geometry of the road or reductions in speed limits.

In our simulations, the infinite road setup consists of a platoon of vehicles. The lead vehicle in that platoon is moving always at a constant speed and is not affected by any added noise or perturbations. 
The road segment with a bottleneck is implemented via (i)~an inflow layer where at each time step vehicles are seeded onto the road segment at a fixed inflow speed and spacing (satisfying an equilibrium relationship) whenever the vehicles already on the road have moved to create enough space, and (ii)~an outflow segment where the speeds of every vehicle that has left the domain is set to a fixed value (lower than the inflow speed). Those vehicles remain being tracked until they can be removed because they have ceased to affect any other vehicles. 
Note that, because below we employ the bottleneck scenario for congested flow, traffic flow theory dictates that it is the outflow conditions that will transport information into the domain. 
We use those different setups in the scenarios of \S\ref{sec:results_and_discussion} as each offers unique insights into traffic behaviour.

\subsection{Traffic Instabilities}
\label{subsec:traffic-instabilities}
The stability of a given equilibrium state can be studied through a linear stability analysis \citep{wilson2011}. 
We linearise the equations of motion around an equilibrium state by choosing the position of vehicle $i$ to be $x_i(t) = (s_{\text{eq}} +\ell)i +v_{\text{eq}}t +y_i$, where $y_i$ is an infinitesimal perturbation. Then, we substitute these vehicle trajectories into \eqref{eq:general-car-following-models}, Taylor-expand around the equilibrium state, and ignore all higher order terms. The perturbation equation obtained is
\begin{align}\label{eq:perturbation_eq}
    \ddot{y}_{i}(t) &= \alpha_1 (y_{i-1} - y_i) - \alpha_2 \dot{y}_{i} + \alpha_3 \dot{y}_{i-1}\;,\\[1em]
    \text{where} \qquad
    \alpha_1 = \frac{\partial f}{\partial s}\;,&\qquad \alpha_2 = \frac{\partial f}{\partial (\Delta v)} - \frac{\partial f}{\partial v} \;,\qquad \alpha_3 = \frac{\partial f}{\partial (\Delta v)}\;,\qquad
\end{align}
and all the partial derivatives are evaluated at the chosen equilibrium state. Then, the growth or decay of solutions to \eqref{eq:perturbation_eq} is characterised by performing a Laplace transform ansatz $y_i(t) = c_i e^{\omega t}$, where $c_i, \omega \in \mathbb{C}$. This yields the interpretation of \eqref{eq:perturbation_eq} as an input/output (I/O) system, $c_i = F(\omega)c_{i-1}$, where the transfer function is
\begin{equation}\label{eq:tranfer_function}
    F(\omega) = \frac{\alpha_1 +\alpha_3 \omega}{\alpha_1 +\alpha_2\omega + \omega^2}\;.
\end{equation}
The temporal growth or decay of the lead vehicle's velocity profile is captured by $\mathrm{Re}(\omega)$ in equation \eqref{eq:tranfer_function}. The frequency of oscillation of the lead vehicle’s velocity profile is represented by $\mathrm{Im}(\omega)$ in equation \eqref{eq:tranfer_function}. The transfer function's modulus $|F|$ is the growth or decay of the perturbation amplitude from one vehicle to the next. An equilibrium state is defined to be string unstable if speed and spacing fluctuations of each vehicle are greater in amplitude than those preceding it \citep{wilson2011}. String instability is a property that can eventually (when growing out of the linear regime) give rise to stop-and-go wave structures in car-following model simulations. With the setup in \eqref{eq:tranfer_function}, stability means that $|F(\omega)|\leq 1 \; \forall \omega \in i\mathbb{R}$ (considering $\mathrm{Re}(\omega) = 0$ is the only way to ensure that $e^{\omega t}$ remains infinitesimally small for all $t\in \mathbb{R}$). 
Evaluating $|F(\omega)|^2$ for $\omega \in i\mathbb{R}$ yields that this stability criterion can be written as a condition on the partial derivatives of $f$, as follows:
\begin{equation}
    \label{eq:stability_condition}
    \alpha_2^2 - \alpha_3^2 - 2\alpha_1 \geq0\;.
\end{equation}
Hence, an equilibrium state's stability is determined by the partial derivatives of $f(s,v,\Delta v)$ with respect to the state variables at that equilibrium state and \eqref{eq:stability_condition}.

\begin{figure}
    \includegraphics[width=\textwidth]{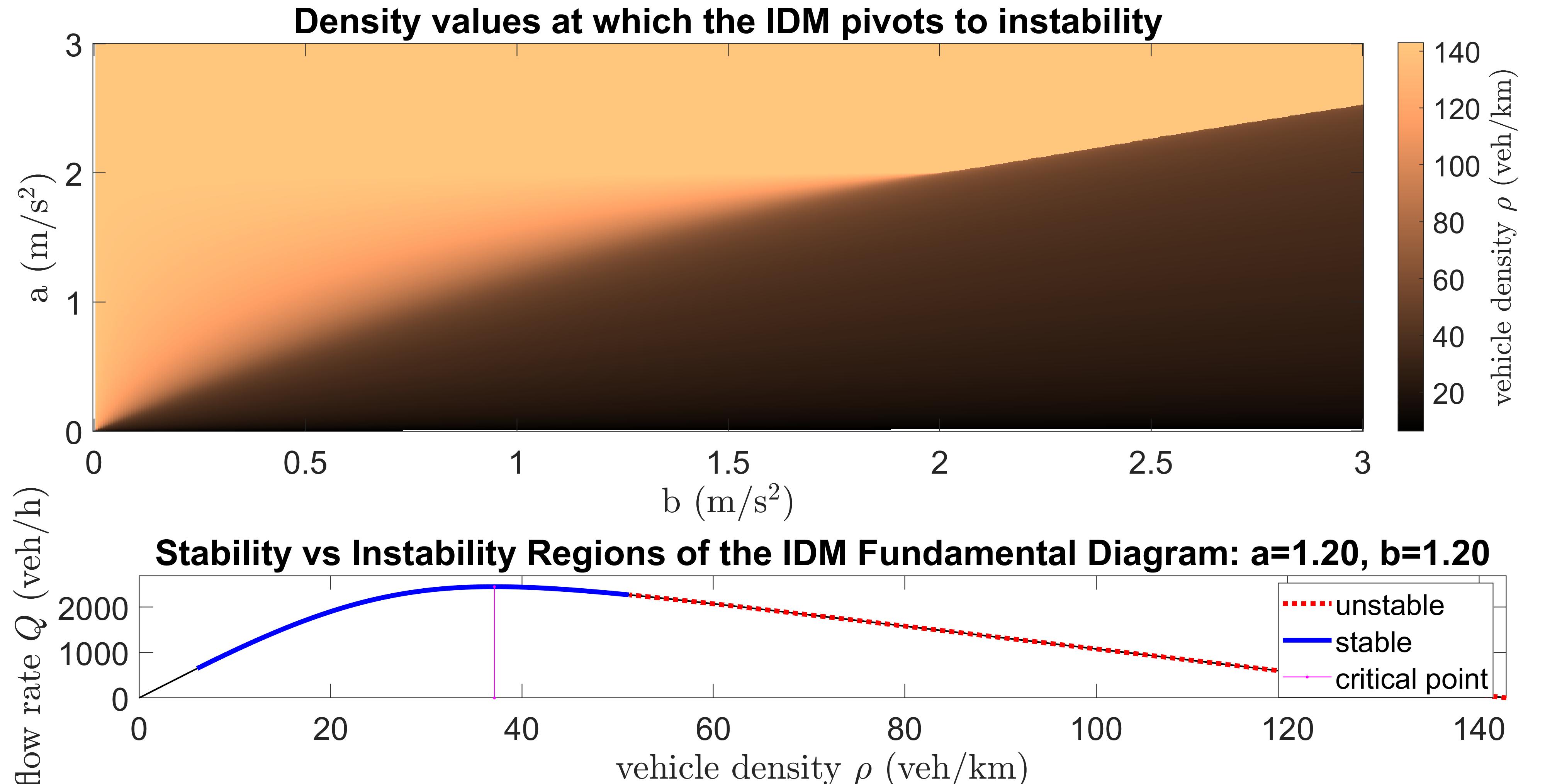}
    \caption{Density value at which the IDM pivots from stable to unstable for a given combination of $a$ and $b$ values (top). It should be noted that a pivoting density value from stable to unstable regimes that is around 140 veh/km means that the model has a tiny region of instability. An example of the visualisation (bottom) of this transition at the level of the IDM fundamental diagram is shown with parameter values $v_0=30\text{m}/\text{s}$, $s_0=2\text{m}$, $\delta = 4$, and $T=1\text{s}^{-1}$ and for specific choices of $a=1.2~\text{m}/\text{s}^2$ and $b=1.2~\text{m}/\text{s}^2$. The blue (solid) part of the fundamental diagram represents the densities of the stable regime, the red (dotted) part represents the densities of the unstable regime (bottom). Clearly, the critical density value at which the highest flow rate occurs is not generally not the same density value at which pivoting to instability occurs.}
    \label{fig:backgrounddensity_vs_a_b}
\end{figure}

For example, some straightforward calculations yield that for the IDM, stability vs.~instability of an equilibrium background density depends only on the choices of the IDM parameters $a$ and $b$. The other parameters determine the shape of the fundamental diagram and are chosen accordingly, where $s_0$ affects the jam density or the maximum traffic road density (reasonable values around $2\text{m}$), $v_0$ affects the maximum flow rate (reasonable values between $30\text{m}/\text{s}$ and $45\text{m}/\text{s}$, but is often set to the maximum speed limit value), $T$ affects the maximum flow rate (reasonable values around $1\text{s}^{-1}$), and $\delta$ affects the curvature of the fundamental diagram (usually set to $4$). Figure~\ref{fig:backgrounddensity_vs_a_b} shows for different choices of $a$ and $b$ the density value at which the IDM pivots from stable to unstable. It also shows an example of the stability vs.~instability regions of the IDM on the level of the fundamental diagram for specific parameter values $a = 1.3\text{m}/\text{s}^2$ and $b = 2\text{m}/\text{s}^2$. It should be noted that when analyzing real-world traffic data the maximum density of vehicles that a road can accommodate while maintaining a relatively smooth and free-flowing traffic condition is often referred to as the ``critical density'' beyond which traffic flow begins to degrade significantly. However, the present analysis of traffic models highlights that the maximum-capacity density and the onset-of-instability density are actually different quantities and may have quite different values. Here, the term ``critical density'' refers to the former.

\subsection{Numerical Methods and Triggering Instabilities}
\label{subsec:numerical-methods}
Simulating the ODE-based microscopic models is carried out using the explicit time-stepping numerical scheme Forward Euler (FE) method. In the absence of noise (see below), one could also use higher order Runge-Kutta methods (e.g., RK4) to gain higher accuracy. In all simulations in this work we choose a time step of size $\Delta t=0.1\text{s}$. This choice ensures that approximation errors induced by the numerical schemes will not create any undesirable effects like vehicle crossings, non-convergence in velocity/position trajectories, or vehicles travelling too far in one time step.

To obtain a simulation of a model that properly captures unstable behaviour, we augment the velocity solution of the model in \eqref{eq:general-car-following-models} by a scaled white noise term (chosen to be a Gaussian with a mean of zero). This means that vehicle trajectories are the solution to a stochastic differential equation arising by augmenting \eqref{eq:general-car-following-models} with an additive Brownian motion. The solution to such stochastic differential equations is a velocity with noise scaled so that the impact of the noise over a time interval is a normally distributed random perturbation with mean zero and variance equal to $\sigma^2$ times the interval’s length, where $\sigma$ denotes the magnitude of the noise. An Euler-Maruyama discretisation of the stochastic differential equation for a discretisation time step $\Delta t$ for the $i^{\text{th}}$ vehicle is given as
\begin{align}
    x_i(t+\Delta t) & = x_i(t) + \Delta t v_i(t)\;, \\
    v_i(t+\Delta t) & = v_i(t) + f(s_i(t), v_i(t), \Delta v_i(t)) + \sqrt{\Delta t}~\mu(0,\sigma)\;.
    \label{eq:car_following_plus_noise}
\end{align}

\begin{figure}
    \includegraphics[width=\textwidth]{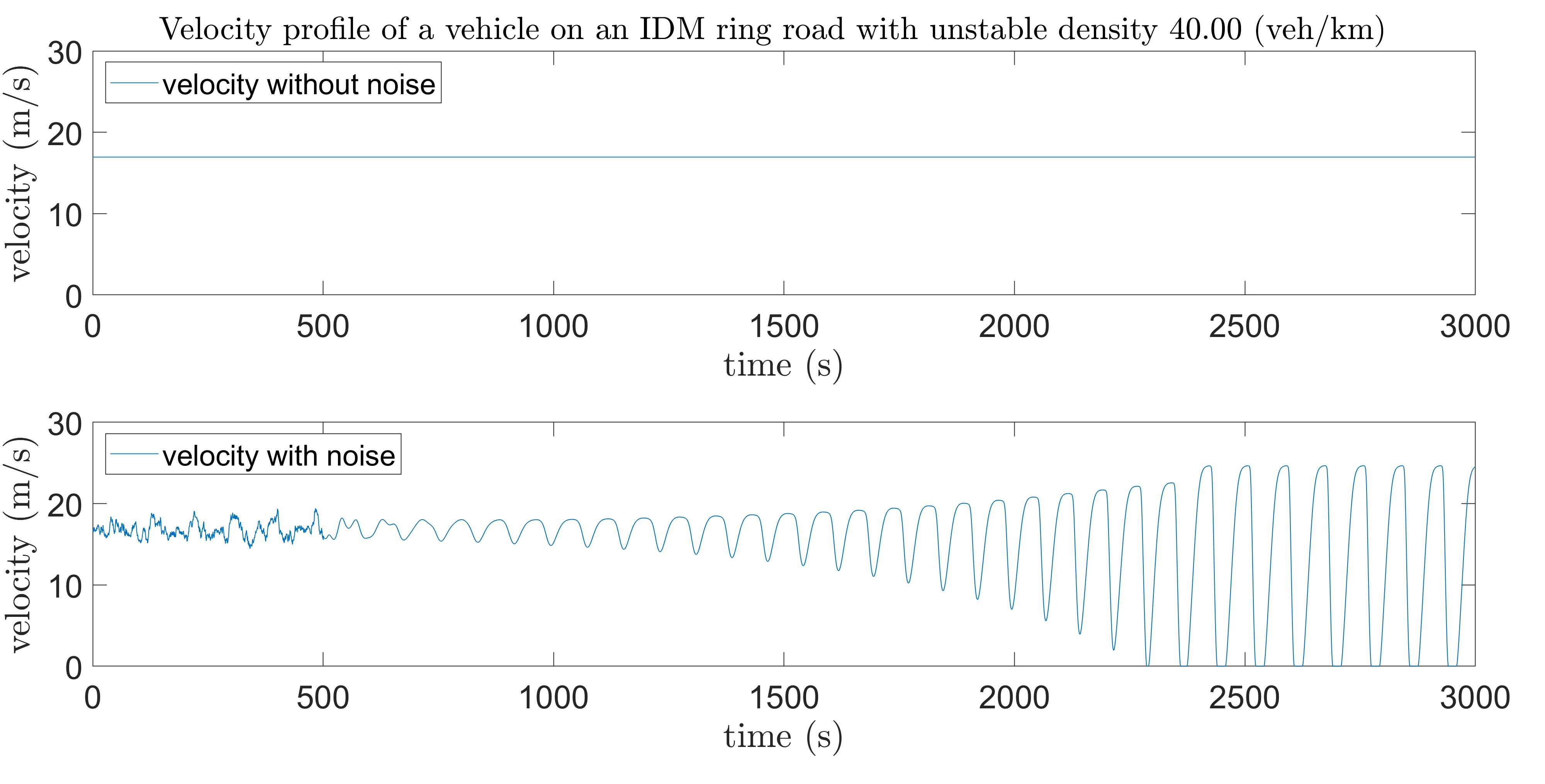}
    \caption{Velocity profile of one vehicle from a $3000\text{s}$ simulation runs of $60$ vehicles on a $1500\text{m}$ IDM ring road initiated at equilibrium with IDM parameters $v_0=30\text{m}/\text{s}$, $s_0=2\text{m}$, $\delta = 4$, $T=1\text{s}^{-1}$, $a=1.3\text{m}/\text{s}^2$, and $b=2\text{m}/\text{s}^2$. The velocity profile (top) from a simulation without noise remains constant, indicating that vehicles remain at equilibrium even though the background density is unstable. The velocity profile (bottom) is from the same simulation but with noise added as described in \eqref{eq:car_following_plus_noise} for the first $500\text{s}$ of the simulation with a magnitude $\sigma = 0.3 \text{m}/\text{s}$.}
    \label{fig:idm_noise_effect}
\end{figure}

The noise in the simulations serves as a simplified model of the inaccuracies that human drivers introduce while trying to maintain a uniform flow driving on a road, or other fast perturbations like gusts of wind. Figure~\ref{fig:idm_noise_effect} shows the importance of including perturbations/noise in simulations for wave development. Two simulations of $60$ vehicles on a $1500\text{m}$ ring-road are initialised at equilibrium, following the IDM with $a = 1.3\text{m}/\text{s}^2$ and $b = 2\text{m}/\text{s}^2$. The background density of this simulation is $40\text{~veh}/\text{km}$, so it lies in the unstable region for the given $a$ and $b$ values (see fundamental diagram in Figure~\ref{fig:backgrounddensity_vs_a_b}), and with enough perturbations, the model dynamics develop waves. The first simulation is carried without noise, and we can see that the speed profile of a vehicle in this simulation remains constant. On the other hand, the second simulation is carried out with the instability triggered by adding noise during the first $500\text{s}$ only, and we see the subsequent development and growth of waves from the variations in the speed profile.

\subsection{Extracting Macroscopic Quantities from Microscopic Models}
\label{subsec:macro-quantities-from-micro-models}
To establish a connection between the micro and macro scales, we derive macroscopic quantities from microscopic models. There are several ways established to do so (like in \citep{LeeLeeKim2001}); one is to use Gaussian kernels of the form 
\begin{equation}
    G(x)=Z^{-1}e^{-(\frac{x}{h})^2}\;,
\end{equation} 
where $h$ is the width of the kernel, chosen to be wider than the minimum spacing between vehicles but small enough to capture localised information about the density and flow rate. For $N$ vehicles on the road, we reconstruct via standard techniques the density as the superposition of these Gaussian kernels
\begin{equation}
    \rho(x,t) = \sum_{j=1}^{N} G(x-x_j(t))\;,
\end{equation} 
and the flow rate as also the superposition of kernels weighted by velocities
\begin{equation}
    q(x,t) = \sum_{j=1}^{N} \dot x_j(t) G(x-x_j(t))\;.
\end{equation}
Consequently we can define the reconstruction of bulk traffic velocity as 
\begin{equation}
    u(x,t) = \frac{q(x,t)}{\rho(x,t)}\;.
\end{equation}

The task of choosing the suitable kernel width $h$ is an art by itself, but 
one should keep in mind that very large $h$ values would smear out features that might be interesting, and very small choices of $h$ would introduce unwanted oscillations.

A crucial advantage of using the above technique to reconstruct macroscopic densities and flow rates is that the reconstructed $(\rho,q)$-pairs exactly satisfy the continuity equation \eqref{LWR-continuity-eq}, and such solutions form lines on the fundamental diagram \citep{KhoudariSeibold}. To prove that latter fact, we consider a solution of the continuity equation \eqref{LWR-continuity-eq} that is a single profile moving with speed $s$ along the road. Then we can write the quantities $\rho$ and $q$ as functions of a single variable $\eta = x-st$. This implies that $\rho_t = -s \frac{d\rho}{d\eta}$ and $q_x = \frac{dq}{d\eta}$, and thus \eqref{LWR-continuity-eq} becomes $\frac{d}{d\eta} (-s\rho+q) = 0$. Integration yields $-s\rho+q = m$, where the integration constant $m$ is the mass flux of vehicles relative to the wave. Thus one obtains the relationship
\begin{equation*}
q = m+s\rho\;,
\end{equation*}
meaning that pairs of density and flow rate, $(\rho,q)$, that satisfy \eqref{LWR-continuity-eq} and are travelling waves, form straight lines in the fundamental diagram plane, where the line's slope $s$ equals the speed of the travelling wave.

\begin{figure}
    \includegraphics[width=\textwidth]{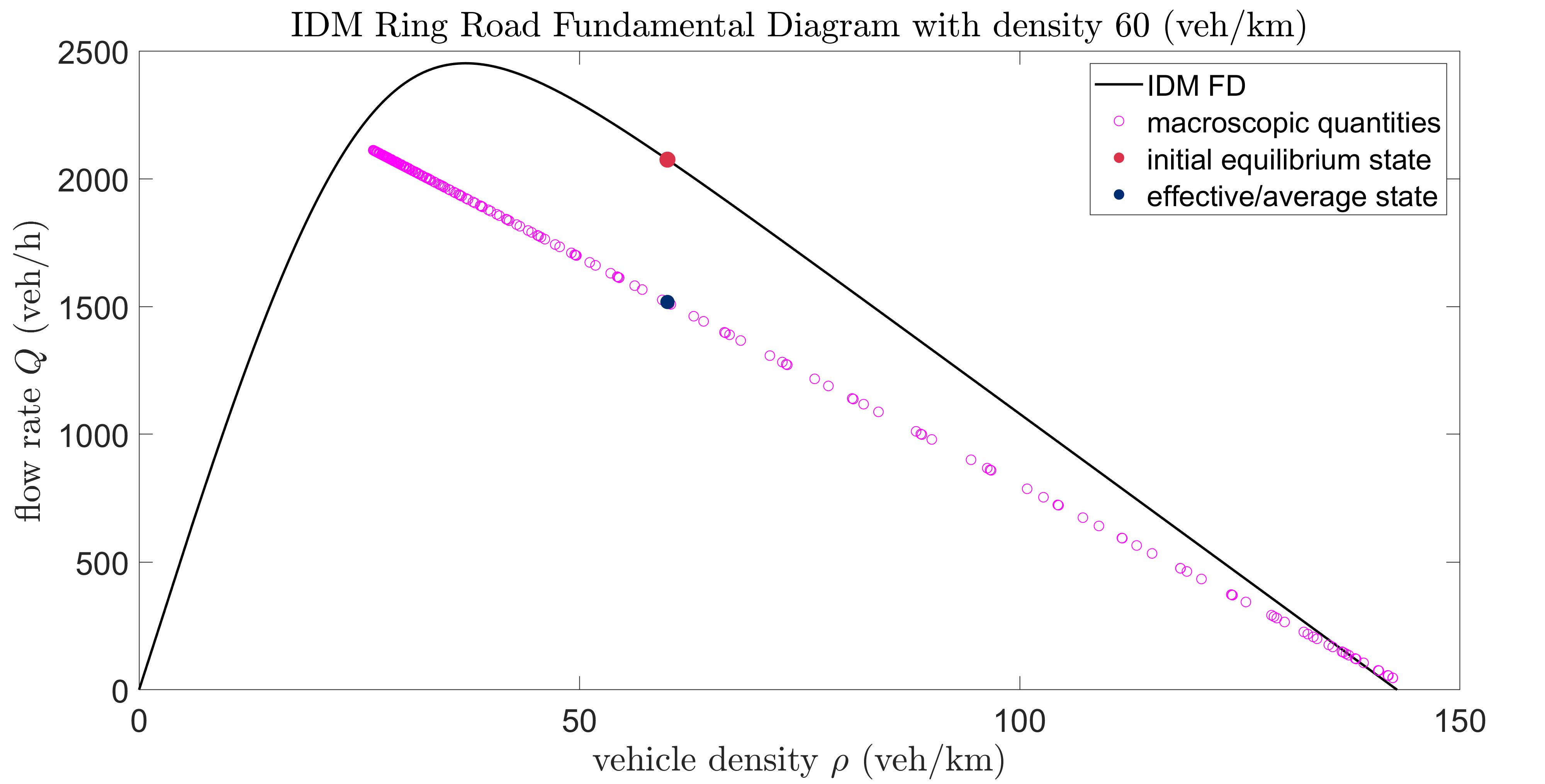}
    \caption{Pairs of the reconstructed macroscopic quantities $(\rho,q)$, as described in \S\ref{subsec:macro-quantities-from-micro-models}, extracted with a kernel width $h=20\text{m}$ at the end time of a $2000\text{s}$ simulation run of $90$ vehicles on a $1500\text{m}$ IDM ring road initiated at equilibrium with IDM parameters $v_0=30\text{m}/\text{s}$, $s_0=2\text{m}$, $\delta = 4$, $T=1\text{s}^{-1}$, $a=1.3\text{m}/\text{s}^2$, and $b=2\text{m}/\text{s}^2$, with noise added for the first $500\text{s}$ of the simulation with a magnitude $\sigma = 0.3 \text{m}/\text{s}$. As predicted by the theory in \S\ref{subsec:macro-quantities-from-micro-models}, all the data points lie clearly on a single line.}
    \label{fig:idm_macroscopic_quantities}
\end{figure}

Therefore, if we plot point-wise all the reconstructed pairs $(\rho,q)$ at a specific instance in time when waves are fully developed they will fall on a line of a slope equal to the speed of the travelling wave (see \S\ref{subsec:wave-characteristics}) and the average of those point-wise (in both time and space) reconstructed $(\rho,q)$-pairs we call the effective or average state (see Figure~\ref{fig:idm_macroscopic_quantities} as an example with the blue dot representing the average state). For a given simulation setup, the curve formed by all the effective states (emerging from many different initial equilibrium states) in the unstable regime is denoted as the \emph{reduced fundamental diagram} curve. Given an initial equilibrium state in the unstable regime, we show interest in determining how the corresponding average state moves in the $(\rho,q)$-space as waves develop in different setups where certain quantities like density, flow rate, and bulk velocity are held fixed (see \S\ref{sec:results_and_discussion}).

\subsection{Wave Characteristics}
\label{subsec:wave-characteristics}
Waves are travelling perturbations in the distribution of vehicles and travelling backwards with respect to vehicles on the road. The state of traffic flow can be determined by the variation of the average vehicle speeds on the road. If this average is high and close to the road's speed limit, then we are in a free flow state. Otherwise, we are in a traffic wave state ranging in intensity from mild (slower vehicles in congestion but with no stoppage) to stop-and-go traffic (experiencing complete stoppage). Quantifying the intensity of traffic waves is affected by the margin of variation of the speeds of vehicles on the road. In this work, we show specific interest in traffic waves and how to interpret them as structures in the fundamental diagram.

\subsection{Traffic Waves as Observed on the Fundamental Diagram}
\label{subsec:waves-in-FD}
Macroscopically, we consider two points in the ($\rho$,$q$)-plane representing the different traffic states: the right state $(\rho_R,q_R)$ and left state $(\rho_L,q_L)$. We define the speed of the shock based on the Rankine–Hugoniot conditions \citep{LeVeque1992} as:
\begin{equation}
    v_s = \frac{q_R - q_L}{\rho_R - \rho_L}\;.
\end{equation}
For a macroscopic interpretation of microscopic waves, we define a `jamiton line' as the nonlinear travelling wave solution on the fundamental diagram \citep{SeiboldFlynnKasimovRosales2013}. The word jamiton was introduced by the authors of \citep{FlynnKasimovNaveRosalesSeibold2009} in analogy to the nonlinear travelling waves in physics called solitons. In the ($\rho$,$q$)-plane, the jamiton line is determined as the best fit line of macroscopic density and flow rate data extracted using Gaussian kernels at a specific instance in time from microscopic simulations. It should be stressed that since we are postprocessing numerical results, the fit is needed because (a)~a travelling wave solution may not be fully established yet; and due to (b)~numerical approximation errors; and (c)~noise effects. The fit will be a straight line segment with slope $v_s$ and two end points $(\rho_R,q_R)$ and $(\rho_L,q_L)$ representing the two traffic states across the wave (see Figure~\ref{fig:idm_macroscopic_quantities} where the endpoints of the line across the fundamental diagram represents the left and right states).

\subsection{Methodology to Compare Jamiton Lines/Wave Speeds for OVM vs.~IDM}
\label{subsec:jamiton_lines_waves_speeds_FD}
With a careful choice of parameters in both the IDM and OVM, the two models can be reasonably compared. For all the comparisons conducted below we use the following parameters: in \eqref{eq:OVM_modified} we set $\alpha=1.085\frac{1}{\text{s}}$, $\beta=22.0779\frac{\text{m}^2}{\text{s}}$, $\nu=2$, $a_\text{m}=1.3\frac{\text{m}}{\text{s}^2}$, $b_\text{m}=5\frac{\text{m}}{\text{s}^2}$, $s_0=2\text{m}$, $v_0=30\frac{\text{m}}{\text{s}}$, whereas in \eqref{eq:IDM_Equation} we set $a=1.3\frac{\text{m}}{\text{s}^2}$, $b=2\frac{\text{m}}{\text{s}^2}$, $v_0=30\frac{\text{m}}{\text{s}}$, $s_0=2\text{m}$, $T=\frac{1}{\text{s}}$, and $\delta=2$. Those parameters are chosen based on the principled methododology explained below.

To be able to compare jamiton lines of two different models, the two models should first have the same fundamental diagrams, i.e., they should exhibit the same equilibrium $(s,v)$-relationships from $f(s,v,0)_{\text{IDM}} = f(s,v,0)_{\text{OVM-Modified}} = 0$.

To match the fundamental diagram of the IDM and OVM we start with an IDM parameter $\delta=2$ (rather than the popular choice $\delta=4$, because for $\delta=2$ the resulting mathematical expressions become simpler) so the IDM equilibrium spacing function can be easily inverted and is given by
\begin{equation}\label{eq:idm_eq_spacing}
    S(v)=\frac{s_0+Tv}{\sqrt{1-(\frac{v}{v_0})^2}}\;.
\end{equation}
From \eqref{eq:idm_eq_spacing} we derive a formula for the OVM optimal velocity function that reads as
\begin{equation}
     V(s) = \frac{-s_0 + \sqrt{s_0^2 - (s_0^2-s^2)(\frac{s^2}{T^2 v_0^2}+1)}}{T (\frac{s^2}{T^2 v_0^2}+1)}\;.
\end{equation}
Note that $V(s)$ is zero for the minimum gap $s_0$ and asymptotes at a maximum velocity $v_0$ for $s\to\infty$.

After achieving the same fundamental diagram equilibrium curves for both models, we now additionally ensure that the stability vs.~instability regions on those curves are also comparable for both models. For that, we start with IDM parameter choices $a$ and $b$ that set the stability regions. We then ensure that the OVM model parameters, namely $\alpha$ and $\beta$, are uniquely determined such that the maximum growth rate of waves and the range of unstable background densities of interest are comparable to those of the IDM. To establish that, we enforce two constraints: 
\begin{enumerate}
    \item The background density at which the corresponding equilibrium solution for both models pivot from stable to unstable is the same. This imposes a linear relation between $\alpha$ and $\beta$ that reads as $\beta=(V'(s_s) - \frac{\alpha}{2})s_s^2$ , where $s_s$ is the equilibrium spacing corresponding to the background density at which the IDM model we started with pivots from stable to unstable.  
    \item The OVM parameter $\alpha$ (and consequently $\beta$) is determined by solving the minimisation problem $$\min_{\alpha}\int_{\Omega_\rho}\lvert \max_\omega F_{IDM}(\rho,\omega)-\max_\omega F_{OVM}(\rho,\omega)\rvert ^2q(\rho)d\rho\;,$$ which minimises the difference of the maximum growth rate of waves $F(w)$ defined in \eqref{eq:tranfer_function} of both models over all possible densities weighted by the flow rate.
\end{enumerate}
After these cross-calibrations of the these two important microscopic models, OVM vs.~IDM, have been conducted, we can now compare them in terms of their jamiton lines generated for different background densities. 

\section{Results and Discussion}
\label{sec:results_and_discussion}
In this section, we present a systematic comparison between the wave properties of the IDM vs.~OVM based on the methodologies discussed above. That is followed by a demonstration of three scenarios based on simulations carried out using the IDM to study the effects of holding certain quantities (density, flow rate, and velocity) fixed on the evolution of the average state. The study is concluded by a discussion on the robustness of the results with a comparison between the implied reduced fundamental diagram shapes for the IDM vs.~OVM.

\subsection{Wave Properties of the IDM vs.~OVM}
Applying the methodologies described above to compare the two models, we find a striking fundamental difference in the waves properties.

\subsubsection{Variation of Jamiton Line Slope vs.~Background Density for the OVM vs.~the IDM}    
\label{subsubsec:slope-vs-density-of-idm-vs-ovm}
\begin{figure}
    \includegraphics[width=\textwidth]{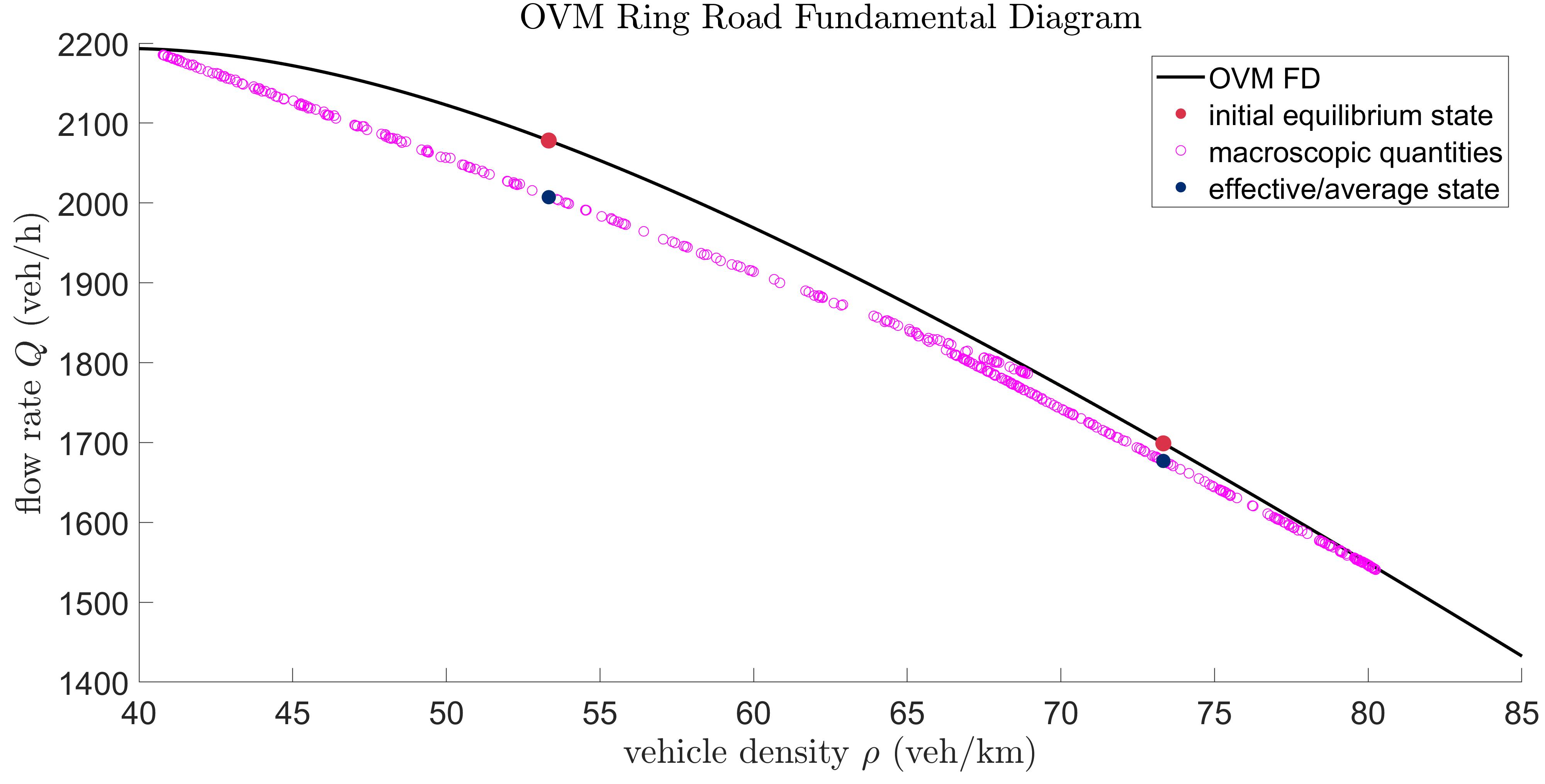}
    \caption{Zoomed plot of the wave solutions of two simulations of the OVM for two different background densities. Wave solutions on the fundamental diagram are represented by the macroscopic reconstruction of the $(\rho,q)$ pairs at the end of the simulation using kernel width $h=20\text{m}$. Simulations are run for $3000\text{s}$ with $80$ vehicles for the first simulation and $110$ vehicles for the second simulation on a $1500\text{m}$ ring road with noise added for the first $400\text{s}$ with magnitude $\sigma=0.04\text{m}/\text{s}$.}
    \label{fig:distinct_jamiton_lines_ovm}
\end{figure}

\begin{figure}
    \includegraphics[width=\textwidth]{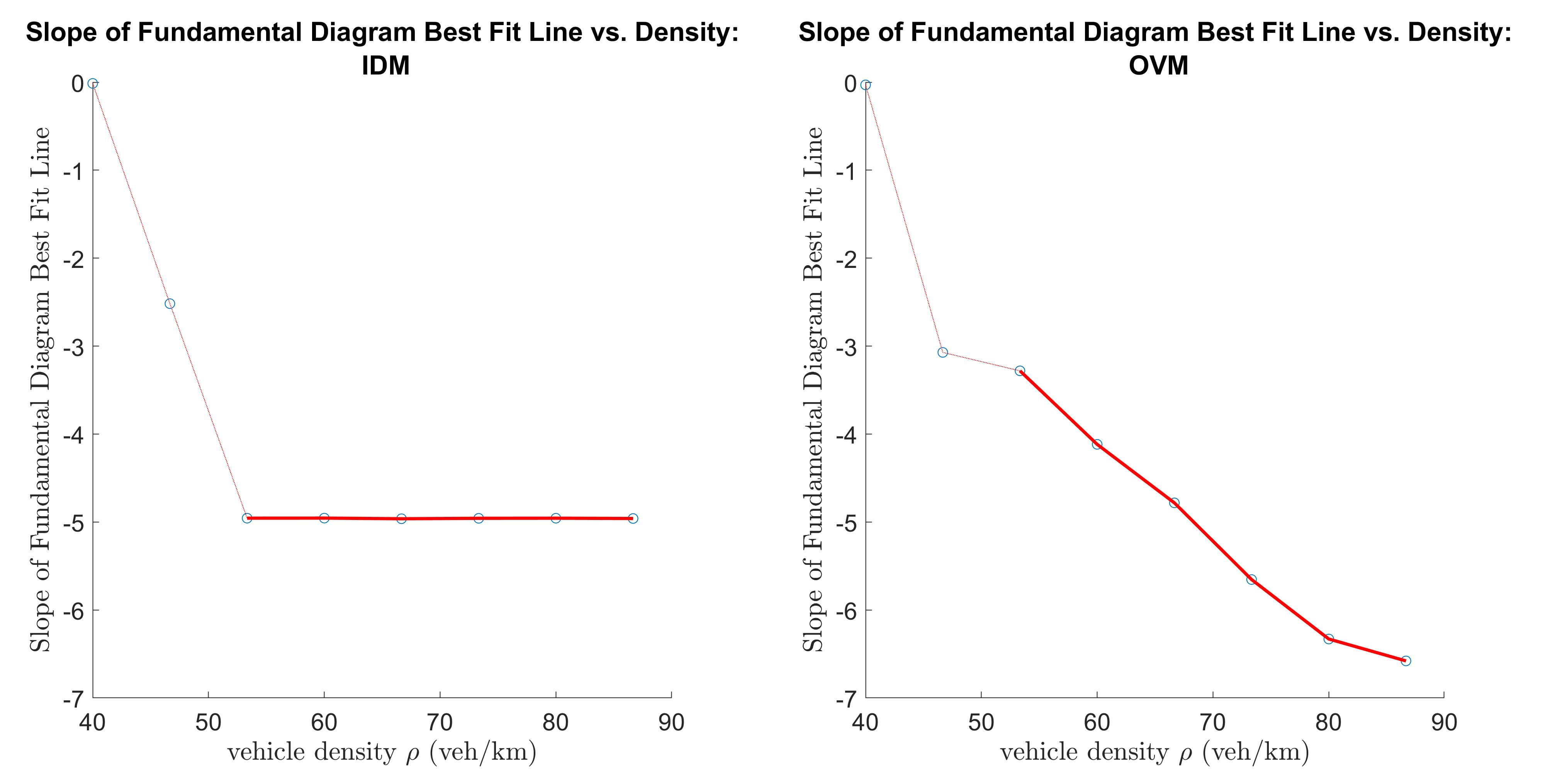}
    \caption{Comparison of the wave speed variation over different background density values for both the IDM and OVM. For each background density, simulations are run on a $1500\text{m}$ ring road for $3000\text{s}$ and the wave speed value is defined to be the average---over a small time interval when waves are fully developed---of slopes of the best fit lines of extracted macroscopic quantities in the $(\rho,q)$-space. Note that for this comparison, the instability threshold is around $40 \text{veh}/\text{km}$, but for densities below $53 \text{veh}/\text{km}$ the instability tends to be not strong enough to yield reliable results.}
    \label{fig:jamiton-slope-vs-density}
\end{figure}

For different background densities the OVM develops different wave solutions shown on the fundamental diagram as different jamiton lines with different wave speeds/slopes (see Figure~\ref{fig:distinct_jamiton_lines_ovm}). In contrast, the IDM has one wave solution developing with the same wave speed (slope of jamiton line) for different background densities that can produce stop-and-go traffic waves (see Figure~\ref{fig:idm_macroscopic_quantities} and ~\ref{fig:jamiton-slope-vs-density}).

To assess the speed of the backwards travelling waves of the OVM as compared to the IDM with stop-and-go waves, an experiment is carried at different background densities for both models. At each background density, multiple simulations were run and the mean of the slopes of the jamiton lines over all simulations was recorded. The simulations were run for 2000s, and data to determine the slope of the jamiton line was collected only during the last 500s to ensure that the traffic waves have fully developed. 

The results of this experiment show that the slope of the jamiton line on the fundamental diagram decreases as a function of background density for the OVM.
In contrast, the slope of the jamiton line remaines constant as a function of background density for the IDM (see Figure~\ref{fig:jamiton-slope-vs-density}).  
The OVM for the densities analysed did develop traffic waves, but it did not produce true stop-and-go waves as the minimum velocity recorded by the vehicles was always greater than $0 \text{km}/\text{hr}$. This is due to the fact that the OVM's solutions are smooth, hence they only approach $v=0$, while the IDM's discontinuous RHS at $v=0$ allows for the solutions to assume patches with exactly $v=0$.
It can be concluded from this experiment that waves travel backward faster at higher densities in the OVM, whereas the speed of the stop-and-go waves in the IDM remains constant for higher densities.  
        
\subsubsection{Wave Strength in the OVM vs.~IDM}
\label{subsubsec:wave-strength-idm-vs-ovm}
Possible ``traffic flow performance'' features that can be considered in quantifying wave development and wave strength are: velocity variation (maximum velocity minus minimum velocity), velocity standard deviation, correlation between velocity of lead and velocity of following vehicle, and energy metrics.

\begin{figure}
    \includegraphics[width=\textwidth]{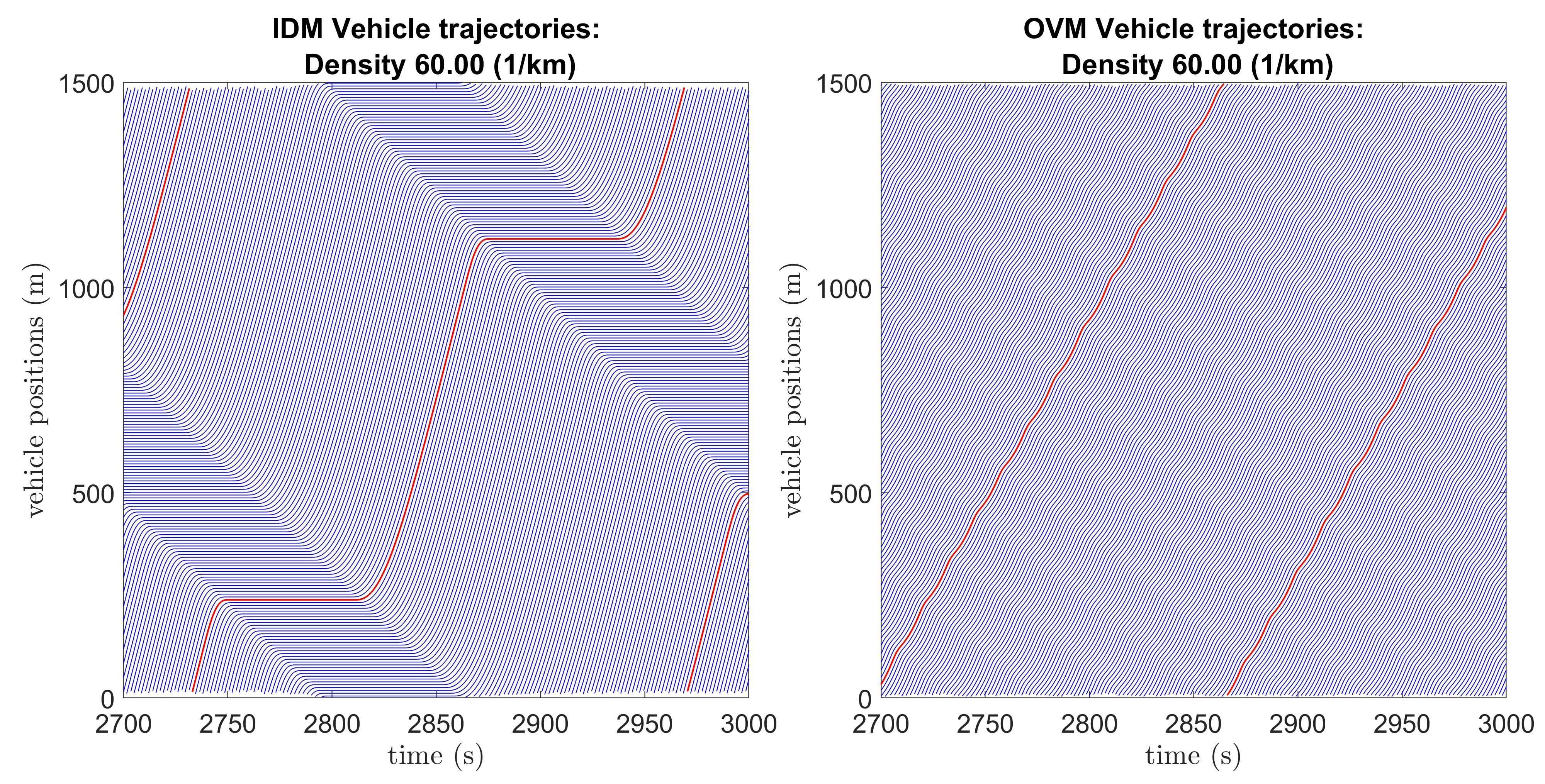}
    \caption{Comparison of vehicle trajectories and wave properties of the IDM vs.~OVM for the same background density $\rho=60 \text{veh}/\text{km}$. Simulations are run on a $1500\text{m}$ ring road for $3000\text{s}$ with noise added only for the first $400\text{s}$ with magnitude $\sigma=0.04 \text{m}/\text{s}$ and trajectories are plotted over the last $300\text{s}$. The trajectory of one vehicle in these simulations is plotted in red for clarity. It should be noted that different choices of parameters/setups will lead to different trajectories, however for comparable OVM and IDM models, the IDM generally tends to result in stronger traffic waves compared to the OVM.}
    \label{fig:trajectory-comparison}
\end{figure}

A simple comparison between the plots of vehicle trajectories resulting from the simulations of the OVM and IDM on a ring road for any chosen unstable background density shows a difference in wave properties. The IDM waves are stronger in terms of the velocity variation (difference between maximum and minimum velocity) and longer in duration than the waves developed by the OVM for the same background density (see Figure~\ref{fig:trajectory-comparison} for an example).  

\begin{figure}
    \includegraphics[width=\textwidth]{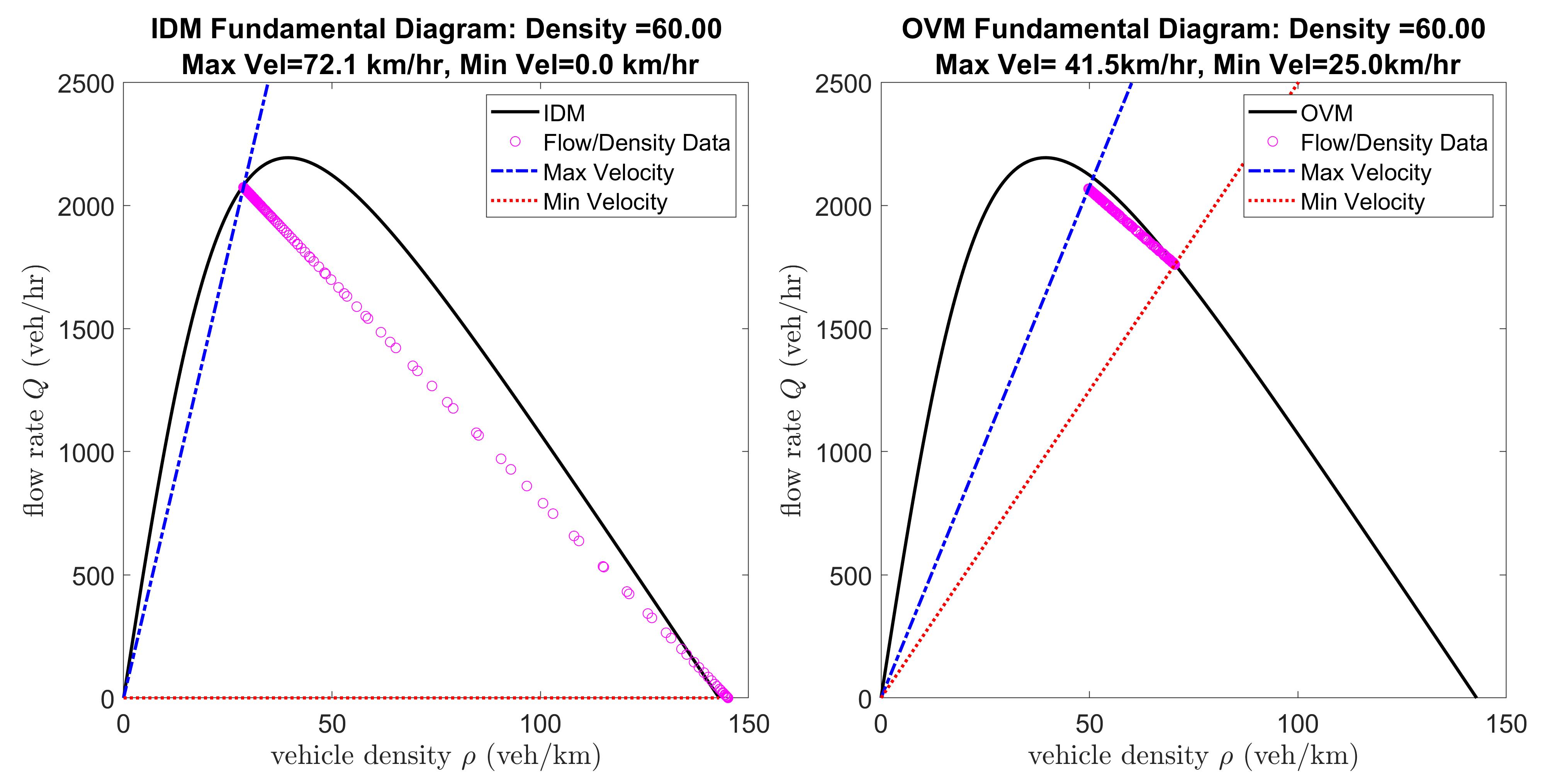}
    \caption{Plots that compare wave strength between the IDM vs.~OVM on the level of the fundamental diagram for the same background density $\rho = 60 \text{veh}/\text{km}$. Simulations are run on a $1500\text{m}$ ring road for $3000\text{s}$ with noise added only for the first $400\text{s}$ with magnitude $\sigma = 0.04 \text{m}/\text{s}$. At the end of the simulations when waves are fully developed, the macroscopic data representing the wave on the fundamental diagram are reconstructed in the $(\rho,q)$-space, and the maximum and minimum velocities in each simulation are recorded. Note that the jamiton lines cross the black equilibrium function.}
    \label{fig:jamiton-comparison}
\end{figure}

A further comparison on the level of the fundamental diagram shows that the OVM jamiton lines are considerably shorter than the IDM jamiton lines and the waves are less strong in terms of velocity variation (see Figure~\ref{fig:jamiton-comparison}). The IDM waves are much stronger, i.e., have a much greater difference between maximum and minimum velocity, which is exemplified by the longer jamiton line on the IDM fundamental diagram when compared to the OVM. The IDM also produces true stop-and-go waves where the minimum velocity reaches zero whereas the OVM minimum velocity is non-zero. This is observed by the jamiton line on the IDM fundamental diagram intercepting with the horizontal axis and by the minimum velocity line coinciding with the horizontal axis.

\subsection{Scenarios}
Considering the fundamental diagram of a second order microscopic car-following model, and starting with a specific equilibrium state on the fundamental diagram where vehicles are equispaced and moving at the same speed, how does the effective (or average) density and flow rate state evolve in the ($\rho$,$q$)-space as we perturb the system?
\\
To answer that we consider three different scenarios where we fix certain quantities (density, velocity, flow rate) and study the effect of fixing such quantities on the evolution of the effective state. 

Note that the presented scenarios are crucial situations that apply in real-world experiments and/or configurations, thus they are also important building blocks for simulations. As highlighted earlier, to isolate the car-following behaviour in this work we only consider a simple case where heterogeneity in driving behaviour and lane switching are not modelled. One can draw an analogy of car-following behaviour in the presence of waves to gas dynamics in higher temperature where more local oscillations rise, and as a consequence the same amount of gas/vehicles ``wiggle" more, thus occupying more space/volume. 
In all three scenarios, we start with equi-spaced vehicles initialised at equilibrium speed and we add perturbations to the system at all times to check how the effective state evolves. Notice that for all three scenarios corresponding to Figures~\ref{fig:Scenario1}, \ref{fig:Scenario2}, and~\ref{fig:Scenario3} (top), as time evolves, the velocities of all vehicles on the road pass through a transient phase before being suitably close to the travelling wave state limit $t\to\infty$ when traffic waves are well developed. Therefore, in all scenarios it is important to consider the the regions in space-time where/when the waves are fully developed, this includes (a)~for all scenarios since we start with equi-spaced configuration, we need to remove the initial and transient layers in time, in addition (b)~for scenarios 2 and 3 we also need to remove spacial boundary layers where waves are not fully developed.

\subsubsection{Scenario 1: Fixed Density}
\label{subsubsec:scenario-1-fixed-density}
\begin{figure}
\includegraphics[width=\linewidth]{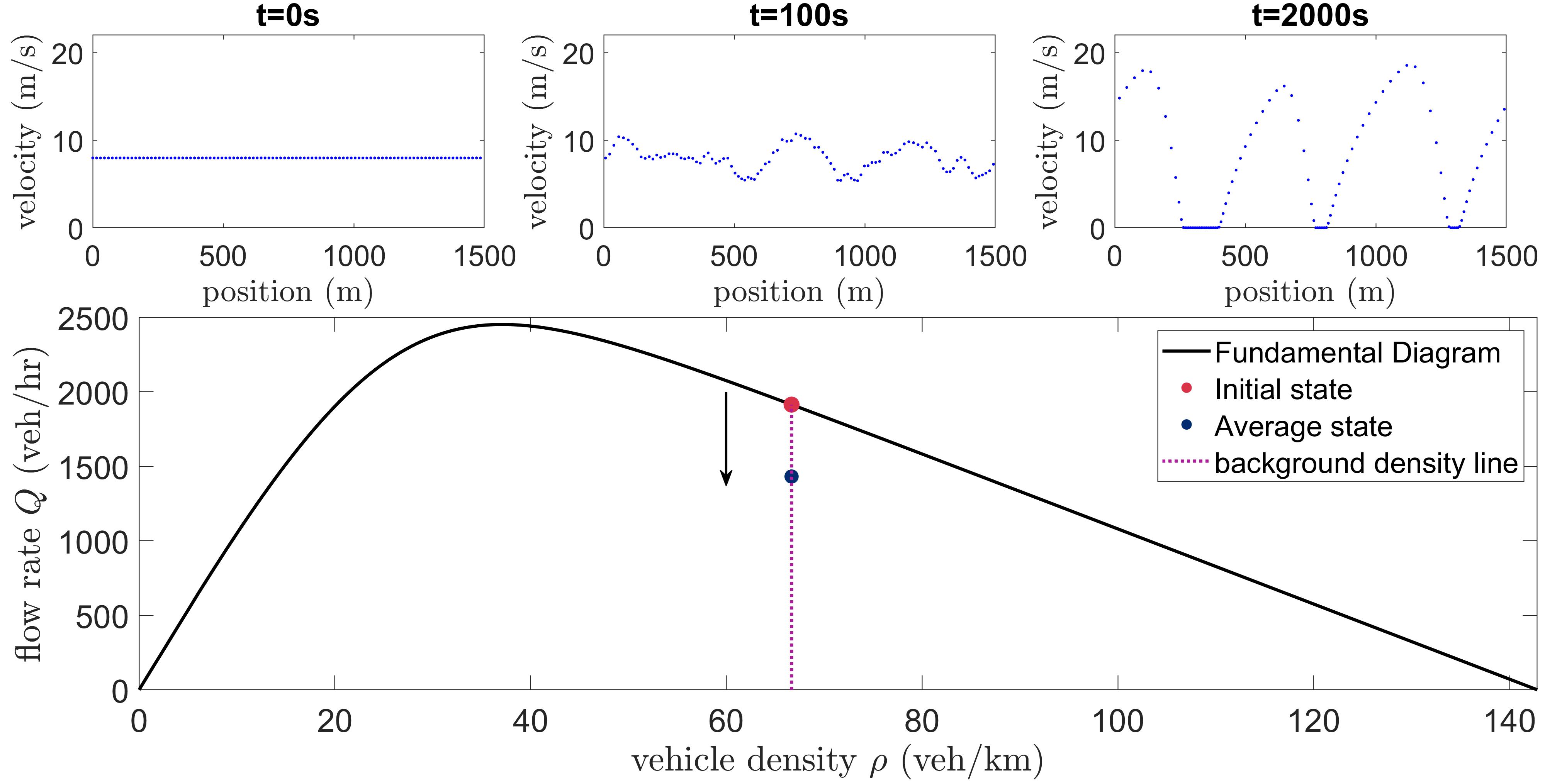}
\caption{The speeds vs.~position (top) of $100$ vehicles on a $1500\text{m}$ IDM ring road simulation at different times. The simulation is run for $2000\text{s}$ and noise is added at all times with magnitude $\sigma=0.3\text{m}/\text{s}$. At $t=0$ all equi-spaced vehicles have the same equilibrium speed, $t=100\text{s}$ lies in a transient phase where the speeds start varying due to the noise and waves start developing, and $t=2000\text{s}$ lies in the wave state limit phase where waves are fully established. On the fundamental diagram (bottom) the corresponding effective/average state starts at the initial equilibrium state and moves downwards on a vertical line as waves develop.}
\label{fig:Scenario1}
\end{figure}

We consider a scenario where the vehicle density remains constant. 
The setup used in this case is a ring road of fixed length and a fixed number of vehicles. Notice in Figure~\ref{fig:Scenario1} that as time evolves, the traffic waves start developing and the effective state of the simulation moves from the initial equilibrium state, vertically downward away from the fundamental diagram, showing a decrease in flow rate while the vehicle density remains constant. It should be noted that the movement line (here: vertical) is due to the fundamental fact that density is conserved. However, the precise effective state location (here: flow rate) and orientation (here: downward) are here only observed.

\subsubsection{Scenario 2: Fixed Speed}
\label{subsubsec:scenario-2-fixed-speed}
\begin{figure}
    \includegraphics[width=\linewidth]{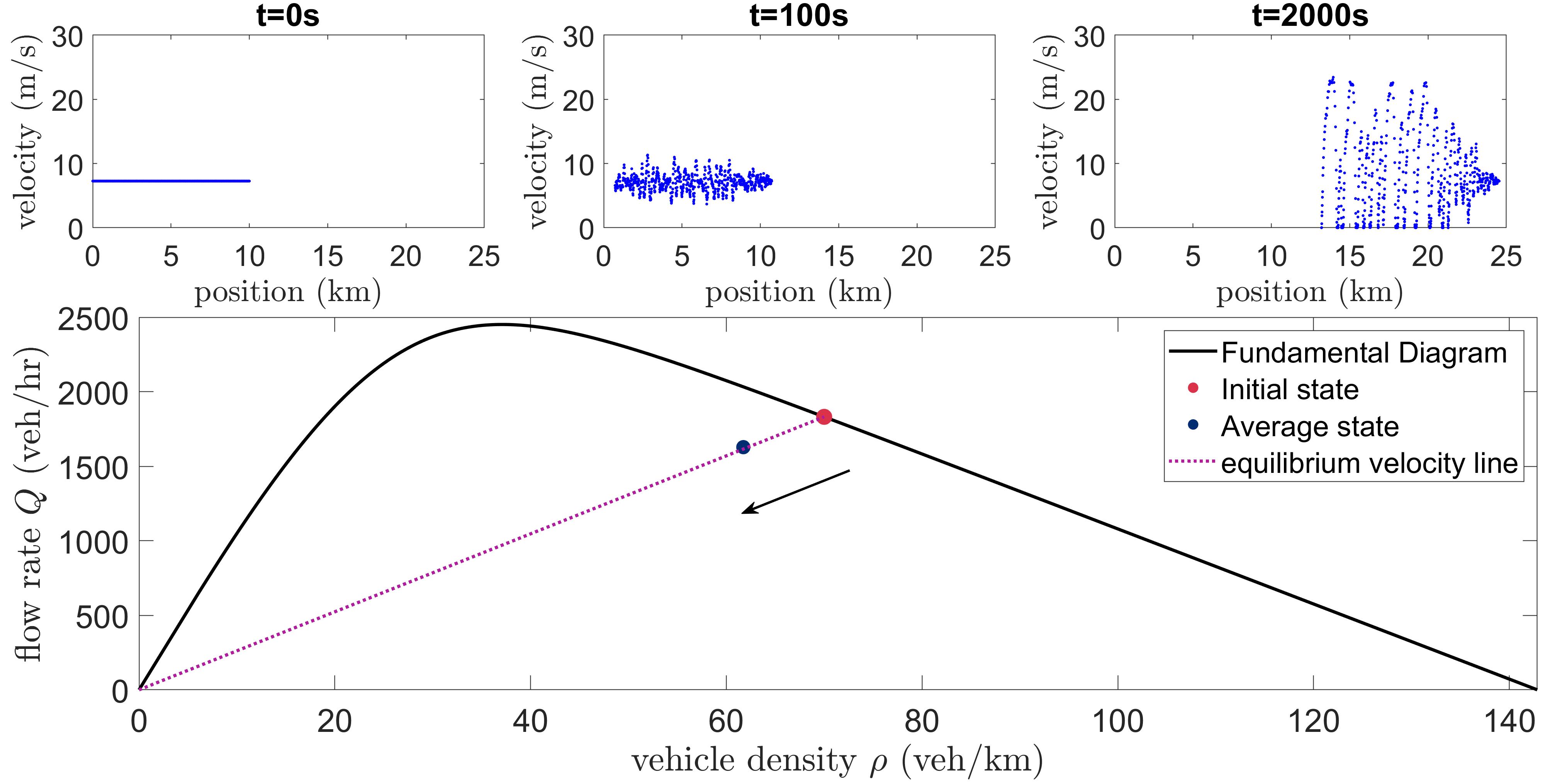}
    \caption{The speeds vs.~position (top) of $700$ vehicles on an infinite road IDM simulation at different times showcasing the initial, transient, and wave phases of the simulation as in Figure~\ref{fig:Scenario1} (top). The simulation is run for $2000\text{s}$, and noise with magnitude $\sigma=0.3\text{m}/\text{s}$ is added at all times to all vehicles excluding the lead. On the fundamental diagram (bottom) the corresponding effective/average state starts at the initial equilibrium state on the fundamental diagram and moves away (towards the origin) on the line passing through the origin as waves develop.}
    \label{fig:Scenario2}
\end{figure}

We consider a scenario where vehicles on average move with a constant speed.  
The setup used in this case is an infinite road with a platoon of vehicles and a lead vehicle moving at the corresponding equilibrium speed at all times. With perturbation, the effective state moves from the initial equilibrium state on a line towards the origin.
The effective state in this scenario is determined without including the layer of vehicles so close to the lead vehicle because waves are not fully established in that layer (this layer depends on the amount of instability but for our simulations we considered a fixed large enough layer of size 100m behind the lead vehicle). Notice that as time evolves, waves start to develop, resulting in a longer road segment occupied between the first and last vehicle compared to the initial platoon length. Consequently the effective density and flow rate proportionally decrease explaining why the effective state moves on the line through the origin (see Figure~\ref{fig:Scenario2}). This verifies that the average speed remains constant. 

\subsubsection{Scenario 3: Fixed Flow Rate}
\label{subsubsec:scenario-3-fixed-flow-rate}
\begin{figure}
    \includegraphics[width=\linewidth]{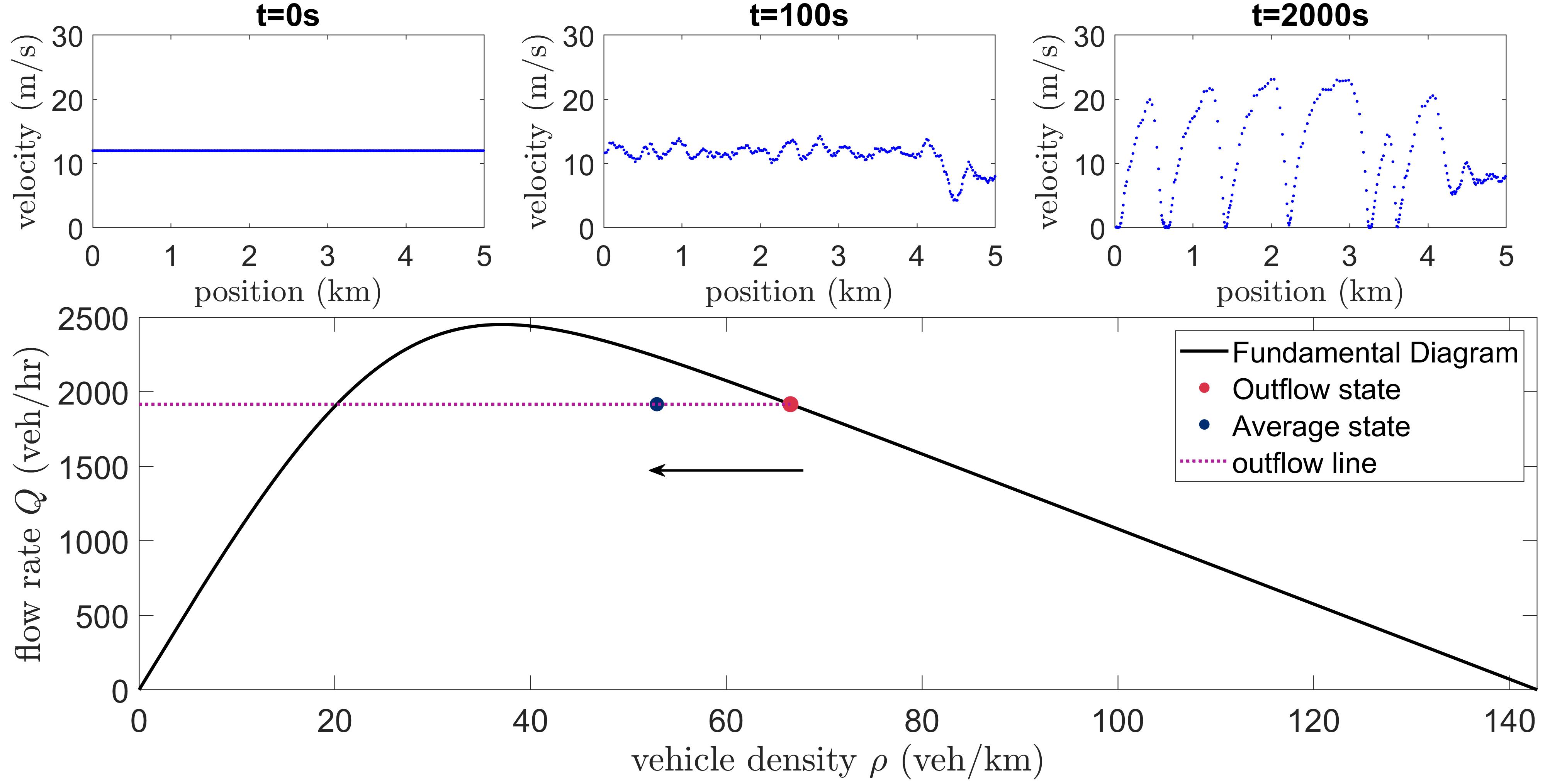}
    \caption{The speeds vs.~position (top) of vehicles on a road segment with a bottleneck for an IDM simulation at different times showcasing the initial, transient, and wave phases of the simulation as in Figure~\ref{fig:Scenario1} (top). The simulation is run for $2000\text{s}$, and noise with magnitude $\sigma=0.3\text{m}/\text{s}$ is added at all times to all vehicles. On the fundamental diagram (bottom) the corresponding effective/average state moves away from the initial equilibrium state and to the left on a horizontal line as waves develop.}
    \label{fig:Scenario3}
\end{figure}

We consider a scenario where on average the flow rate is fixed. The setup used in this case is a bottleneck scenario where we consider a road segment with limits on the outflow rates and inflow conditions (as described in \S\ref{subsec:setups}).
With perturbation, the effective state moves to the left on a horizontal line, away from the equilibrium state. The effective state in this scenario is determined without considering the the inflow and outflow layers to avoid boundary effects in those regions (in our simulations we considered a fixed large enough layer of size 500m near the inflow and outflow regions). As time evolves, a congested region (a traffic jam) starts to form and grow into the free flow region upstream of the bottleneck. Perturbations will trigger the development of traffic waves in the congested region and consequently the effective state of traffic on that segment moves to the left on a horizontal line on the fundamental diagram (see Figure~\ref{fig:Scenario3}). 

The average speed increases on this road segment because vehicles are prevented from entering the road segment as the traffic jam moves further upstream. This explains the inverse proportionality: a decrease in average density and an increase in average speed consequently resulting in a constant effective flow rate. At first glance, it may appear counterintuitive why the presence of waves, which causes a \emph{lower} effective density and \emph{higher} speed, is undesirable. However, the reason is that the waves allow to fit \emph{fewer} vehicles onto the considered road segment, thus preventing more vehicles from advancing forward along the road. 

Those three scenarios are considered the building blocks for more complicated and thus more realistic scenarios. 

\begin{figure}
    \includegraphics[width=\linewidth]{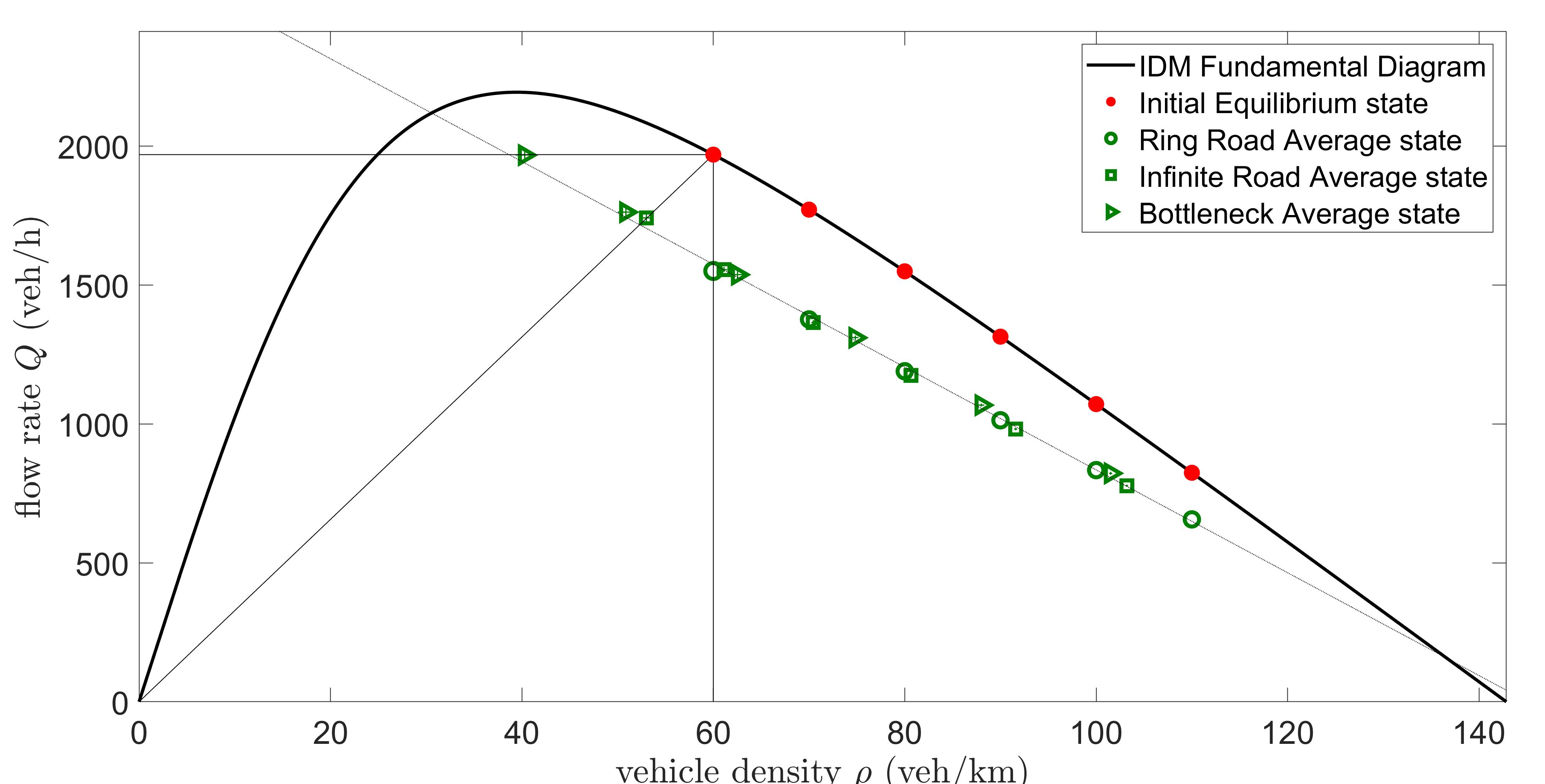}
    \caption{The average of effective states with error bars in both the density and the flow rate direction over an ensemble of 25 IDM simulations, per scenario, per background density in the unstable regime. IDM simulations are run for a long time to ensure that waves are fully developed in the presence of added noise. For each background density and for each scenario, the effective state for each simulation is calculated as the average of the extracted density and flow rate quantities over a final time interval. The plotted averages of effective states are computed as the mean of the effective states over all ensemble simulations, and the error bars represent the standard deviations. This plot shows that for the IDM the shape of the reduced fundamental diagram is scenario-independent as all effective states for all scenarios lie on the same line.}
    \label{fig:robustness}
\end{figure}

\begin{figure}
 \centering
    \includegraphics[width=\linewidth]{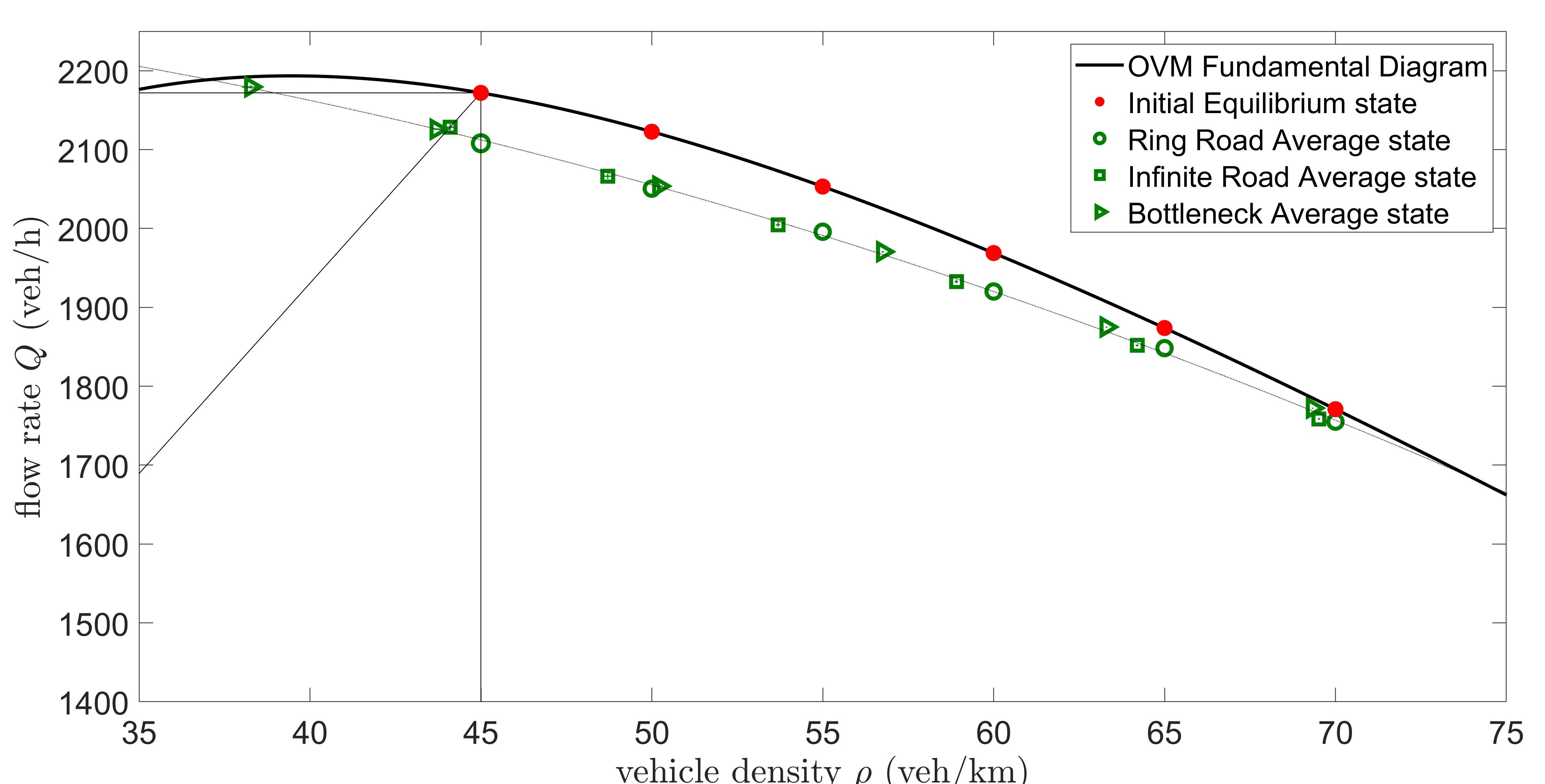}
    \caption{The average of effective states with error bars in both the density and the flow rate direction over an ensemble of 25 OVM simulations, per scenario, per background density in the unstable regime. OVM simulations are run for a long time to ensure that waves are fully developed in the presence of added noise. For each background density and for each scenario, the effective state for each simulation is calculated as the average of the extracted density and flow rate quantities over a final time interval. The plotted averages of effective states are computed as the mean of the effective states over all ensemble simulations, and the error bars represent the standard deviations. This plot shows that for the OVM the shape of the reduced fundamental diagram is scenario-independent as the effective states for all scenarios lie on the same fitted jamiton curve.}
    \label{fig:robustness_ovm}
\end{figure}

\subsection{Robustness and Consistency Across Scenarios}
\label{subsec:robustness}
In this subsection, we (i)~ensure the robustness of the process of finding the effective/average state and its evolution in the density-flow rate plane for each scenario, (ii)~determine the curves that the various average states form, and (iii)~compare the effective states curves resulting in the reduced fundamental diagram for the IDM vs.~OVM in all three scenarios. We run an ensemble of simulations at each background density of interest with the initial equilibrium states chosen to be in the unstable regime and the density is not extremely high (i.e., not close to the maximum density). This ensures that waves will develop, and eliminates non-interesting cases of fully occupied roads. For each background density, the mean of the calculated average states from an ensemble of 25 simulations is plotted for the IDM and OVM in Figures~\ref{fig:robustness} and~\ref{fig:robustness_ovm} respectively, with error bars in both the density and flow rate direction representing the standard deviations over the respective ensembles. The same methodology is carried for all three scenarios at each initial equilibrium state.

The results verify that for both the IDM and OVM, the effective states start at the initial equilibrium state and move away from the fundamental diagram: (i)~downwards on a vertical line in the ring road scenario, (ii)~towards the origin on a line connecting the initial equilibrium state with the origin in the infinite road scenario, and (iii)~to the left on a horizontal line in the bottleneck scenario, for all unstable background densities. This confirms that although each simulation has a random component, in all three scenarios we have the same fundamental, systematic principle: waves correspond to the same kind of car-following behaviour that the vehicles exhibit on the road, independent of the scenario set up; it is only the manifestations where the corresponding effective states end up in the ($\rho$,$q$)-space that is different depending on the setup.

A further important observation can be made about the shape of the reduced fundamental diagram for the IDM vs.~OVM. Note that for the IDM the averaged effective states for all unstable background densities and from all three different scenarios are reasonably close to one single line (a minor exception is for lower background densities that are close to the boundary of instability), verifying that the shape of the reduced fundamental diagram is scenario-independent as all effective states for all scenarios in the unstable regime lie on the same line (see Figure~\ref{fig:robustness}). For the OVM, the averaged effective states for all unstable background densities and from all three different scenarios are reasonably close to a fitted jamiton \emph{curve} rather than a line (see Figure~\ref{fig:robustness_ovm}). This also verifies that the shape of the reduced OVM fundamental diagram is scenario-independent as all effective states for all scenarios in the unstable regime lie on the same curve. The results in this study show that the presence of traffic waves always yields the same effective velocity-spacing relationship, thus microscopic models in the unstable flow regime exhibit many of the fundamental structural properties that have been proven for second-order macroscopic traffic models \citep{SeiboldFlynnKasimovRosales2013, RamadanRosalesSeibold2019}.

\section{Conclusions and Outlook}
\label{sec:conclusions_outlook}
Traffic models are widely used in multiple scales, most prominently microscopic and macroscopic. Research works and practical applications usually employ either one or the other scale. In this work we connected the micro scale to the macro scale by studying the macroscopic interpretation of microscopic waves in car-following models. Complementing the study of those waves in micro-simulations, data analysis tools, and mathematical models, this work establishes a framework to study the models in a systematic hierarchy of tests that isolate the car-following dynamics and structurally connect the microscopic vehicle scale with the meaningful macroscopic effective flow quantities. This framework is based on three simple building block scenarios where one of the three specific quantities density, velocity, and flow rate is held fixed to study the amplification of waves under small perturbations. Results are compared for the two models IDM and OVM, where the two models are calibrated to have the same fundamental diagram and similar unstable regimes. By carrying out simulations using the two models, the difference in wave speeds and jamiton lines on the fundamental diagram between the two models and the different criteria for determining the strength of waves are highlighted. It should be noted that the waves exhibit the same kind of characteristic behaviour independent of the scenario used, with a difference in wave strength depending on the model used. However, the way the average state of a simulation (in the presence of waves) moves in the ($\rho$,$q$)-space depends on the setup of the scenario, and consequently determines the shape of the corresponding reduced fundamental diagram. For the IDM, the reduced fundamental diagram is scenario-independent with all the averaged effective states from all three scenarios lying on the same line, whereas the reduced fundamental diagram of the OVM is scenario-independent with the averaged effective states from all three scenarios lying on a jamiton curve. Looking forward, we aim to establish fundamental insights into the performance of traffic control strategies on sparse Connected and Automated Vehicles (CAVs) that aim at smoothing the flow and dampening the waves. The principle feasibility of the new paradigm of Lagrangian flow smoothing via a few CAVs has been in simulation \citep{rajamani2002,talebpour2015, CuiSeiboldSternWork2017} as well as experimentally \citep{SternCuiDelleMonacheBhadaniBuntingChurchillHamiltonHaulcyPohlmannWuPiccoliSeiboldSprinkleWork2018, WuSternCuiDelleMonacheBhadaniBuntingChurchillHamiltonHaulcyPiccoliSeiboldSprinkleWork2018,AZExperimentData2017}. However, in all those situations the consequences of the CAV-based controls are empirical in nature, i.e., they are \emph{observed} from real-world experiments or from micro-simulations, rather than resulting from the principled analysis. Thus, a natural generalisation of the present work is to aim to establish the macroscopic manifestations of heterogeneous micro-scale behaviour in a systematic, principled fashion.


\section{Acknowledgements}
This material is based upon work supported by the U.S.\ Department of Energy’s Office of Energy Efficiency and Renewable Energy (EERE) under the Vehicle Technologies Office award number CID DE--EE0008872. The views expressed herein do not necessarily represent the views of the U.S.\ Department of Energy or the United States Government.
Research was sponsored by the DEVCOM Analysis Center and was accomplished under Cooperative Agreement Number W911NF-22-2-0001. The views and conclusions contained in this document are those of the authors and should not be interpreted as representing the official policies, either expressed or implied, of the Army Research Office or the U.S. Government. The U.S. Government is authorised to reproduce and distribute reprints for Government purposes notwithstanding any copyright notation herein.

\section{Biographical note}
\textit{Nour Khoudari} (\orcidicon{0000-0002-9987-6525} 
\href{https://orcid.org/0000-0002-9987-6525}{orcid.org/0000-0002-9987-6525}, \texttt{nour.khoudari@temple.edu}) is a Ph.D.\ candidate in mathematics at Temple University. Her research focus is in applied mathematics, particularly traffic modeling, multi-agent systems, optimization, and control.

\textit{Rabie Ramadan} received his Ph.D.\ degree in mathematics from Temple University. He is a Lead Data Scientist with Legal \& General America, Frederick, Maryland. His focus is on optimization and machine learning.

\textit{Megan Ross} received her B.E.\ degree in Electrical and Computer engineering from Temple University. She is a Data Engineer at Comcast Philadelphia, Pennsylvania. 

\textit{Benjamin Seibold}(\orcidicon{0000-0003-2879-6402} 
\href{https://orcid.org/0000-0003-2879-6402}{orcid.org/0000-0003-2879-6402}, \texttt{seibold@temple.edu}) is a Professor of Mathematics and Physics, and the Director of the Center for Computational Mathematics and Modeling, at Temple University. His research areas, funded by NSF, DOE, DAC, USACE, USDA, and PDA, are computational mathematics (high-order methods for differential equations, CFD, molecular dynamics) and applied mathematics and modeling (traffic flow, invasive species, many-agent systems, radiative transfer).

\bibliographystyle{plain}
\bibliography{micro_macro_references}

\end{document}